# Past Psychedelic Use Predicts Divergent Thinking

Gregory J Pope[1*], Christopher Timmermann[2*], William Trender[3], Peter J Hellyer[4], Maria Bălăeț[3,4*], Ruben Eero Laukkonen[5,6]

[1] Southern Cross University, Faculty of Health, Gold Coast, QLD, Australia
[2] UCL Centre for Consciousness Research, Experimental Psychology, University College London, London, UK
[3] Department of Brain Sciences, Imperial College London;
[4] Department of Neuroimaging, IoPPN, King's College London,
[5] Flourishing Intelligence Program, Linacre College, University of Oxford
[6] Centre for Eudaimonia and Human Flourishing, Linacre College, University of Oxford

*For correspondence: gregoryjamespope@gmail.com; m.balaet17@imperial.ac.uk

**Abstract**

Psychedelics have shown potential in treating a range of mental health conditions, yet far less is known about their impact on creativity. This study examined three components of creativity—divergent thinking, cognitive reflection, and insight—in a large sample (N = 5,905) from the Great British Intelligence Test. We compared performance between individuals with past psychedelic use and those without such use. Psychedelic users scored significantly higher on divergent thinking than both non-drug users and drug users who had not used psychedelics. However, they did not score higher on measures of cognitive reflection, number of insights, or insight accuracy. These findings suggest that naturalistic psychedelic use may be associated with enhanced divergent thinking, but not enhanced insight-related performance. Future research should aim to establish causality through prospective designs and controlled studies incorporating long-term follow-up, biological data, and personality structure assessment.

**Funding:** During the time this research project was carried out MB was supported by the Medical Research Council Doctorate Training Programme at Imperial College London and WT was supported by the EPSRC Centre for Doctoral Training in Neurotechnology. PJH is, in part, supported by the National Institute for Health Research (NIHR) Biomedical Research Centre at South London and Maudsley NHS Foundation Trust and King's College London. **Conflict of interest:** PJH is co-owner and director of H2 Cognitive Designs LTD which support online studies and develop custom cognitive assessment software and reports personal fees from H2 Cognitive Designs LTD outside the submitted work. The remaining authors declare that the research was conducted in the absence of any commercial or financial relationships that could be construed as a potential conflict of interest. WT is employed by H2 Cognitive Designs LTD.

**Introduction**

"Taking LSD was a profound experience, one of the most important things in my life." - Steve Jobs (Isaacson, 2011, p. 41)

Creativity is pivotal not only for developing novel and practical solutions to problems, but also for navigating the complexities of a rapidly changing world. It is central to a wide range of domains and contributes to scientific discovery, economic development, and social change (Aylesworth, 2013). Many artists, scientists, and entrepreneurs, including Aldous Huxley and Steve Jobs, have described psychedelics as supporting or enhancing creative insight (Wießner et al., 2022). Historical accounts further suggest that psychedelic experiences have influenced developments in fields such as chemistry, pharmacology, mathematics, theoretical physics, computing, and software development (Gandy et al., 2022), as well as art, music, and literature (Baggott, 2015). Given creativity's importance for adaptation and problem-solving, exploring factors that may enhance it is warranted. The present study tests the hypothesis that past psychedelic use is positively associated with several aspects of the creative process, including divergent thinking, cognitive reflection, and insight.

**Psychedelics and creativity**

Following a prolonged lull caused by widespread prohibition and negative public discourse (Bălăeț, 2024; Bălăeț et al., 2024), psychedelic substances have experienced a resurgence in research interest (Belouin & Henningfield, 2018). This renewed attention reflects growing evidence that psychedelics may offer therapeutic benefits across a range of mental health conditions (Carhart-Harris et al., 2016; Gasser et al., 2014; Krediet et al., 2020; De Veen et al., 2017; Bogenschutz et al., 2015; Johnson et al., 2017; Foldi et al., 2020; Moreno et al., 2006; Flanagan & Nichols, 2018), with some studies reporting long-lasting improvements from even a single dose (Knudsen, 2023). These effects have also been documented in naturalistic settings (Aday et al., 2020; Bălăeț et al., 2025). Their psychological impact is thought to arise primarily from agonism at the 5-HT2A serotonin receptor and altered blood flow and network dynamics in regions such as the default mode network (Carhart-Harris & Friston, 2019), processes that have been linked to rapid shifts in belief and cognition.

Recent years have also seen renewed interest in the idea that psychedelics may promote creativity (Wießner et al., 2022). Psychedelic use has been associated with the personality trait of openness to experience (Erritzoe et al., 2019), which itself is linked to creativity (McCrae, 1987). However, few studies have directly examined psychedelic-related changes in creativity using cognitive tasks, and the limited available evidence has been inconsistent (Bălăeț, 2022). One study on psilocybin suggested that although participants *felt* more insightful and creative during the acute state, their objective performance did not differ from baseline; only one week later did participants generate a greater number of novel ideas (Mason et al., 2021). The vividness and novelty of psychedelic experiences may therefore be mistaken for genuine creative output, and measuring objective creativity remains complex (Baggott, 2015). Although unconstrained thought flow has been proposed as a mechanism through which psychedelics disrupt maladaptive thinking patterns (Carhart-Harris & Friston, 2019), the discrepancy between subjective impressions and objectively measured outcomes highlights the need to distinguish perceived creativity from demonstrable creative performance.

**Psychedelics, Divergent Thinking, Cognitive Reflection, and Insight**

One definition of creativity is the ability to produce original and valued ideas, acts, or objects (Csikszentmihalyi, 1997). This capacity involves at least three key cognitive processes: divergent thinking, cognitive reflection, and insight. Divergent thinking supports the generation of novel ideas (Addis et al., 2016), cognitive reflection enables deliberate and effortful reasoning to refine those ideas (Erceg et al., 2020), and insight involves sudden cognitive shifts that give rise to novel realisations that feel immediately true (Kounios & Beeman, 2014; Laukkonen et al., 2023). Together, these processes contribute to individual problem-solving and to broader advancements in science, art, and innovation (Aylesworth, 2013). Recent work also suggests that insight experiences may play a key role in the therapeutic effects of psychedelics (Kugel et al., 2024; see also Tulver et al., 2023).

Divergent thinking is associated with imagination and future event simulation (Guilford, 1967; Addis et al., 2016) and is commonly assessed using tasks such as the Alternative Uses Task, the Torrance Test of Creative Thinking (TTCT), and the Divergent Association Task (DAT) (Guilford, 1967; Torrance, 1966; Olson et al., 2021). Unlike divergent thinking, convergent thinking refers to the ability to reach a well-defined solution and can help determine the usefulness of ideas generated during divergent thinking (Cropley, 2006). Early studies on LSD and mescaline suggested psychedelics might enhance divergent thinking, but methodological limitations, such as small samples, inadequate controls, and expectancy effects, restrict the conclusions that can be drawn (Baggott, 2015; Harman et al., 1966; McGlothlin et al., 1967; Zegans et al., 1967). More recent studies of psilocybin and ayahuasca show mixed findings: some report short-term increases in divergent thinking and delayed improvements in convergent thinking, whereas others observe reductions during acute intoxication (Frecska et al., 2012; Kuypers et al., 2016; Mason et al., 2019, 2021). Microdosing research has also reported possible cognitive benefits, though placebo effects may explain much of the perceived improvement (Prochazkova et al., 2018; Szigeti et al., 2021). Overall, results remain inconclusive, partly due to variation in psychological traits (set), physical and social environment (setting), and study designs, with limited work examining long-term outcomes.

Cognitive reflection is the ability to override intuitive responses in favour of deliberate, analytic reasoning and is central to decision-making and problem-solving (Bialek & Pennycook, 2018; Erceg et al., 2020). It is typically conceptualised within dual-process theory, where System 1 processes are fast and intuitive, and System 2 processes, crucial for cognitive reflection, are slower and effortful (Evans & Stanovich, 2013; Patel et al., 2019).

The Cognitive Reflection Task (CRT) measures this ability and predicts a range of outcomes, including reasoning quality, susceptibility to pseudo-profound statements, and certain life decisions (Frederick, 2005; Erceg et al., 2023). Although no studies have directly tested how psychedelics affect cognitive reflection, research suggests they may enhance psychological flexibility (Davis et al., 2020) and reduce functional fixedness (Sweat et al., 2016), potentially supporting more reflective thinking. However, psychedelics have also been linked to increased suggestibility (Carhart-Harris et al., 2015), socially-driven belief adoption (Timmermann et al., 2021), and false memories (Timmermann, 2022), raising the possibility that cognitive reflection could be diminished under certain conditions.

Psychedelics often induce profound insight experiences marked by sudden cognitive shifts and an emotional "Aha!" moment (Davis et al., 2021; Öllinger & Knoblich, 2009; Webb et al., 2018; Laukkonen & Tangen, 2017). Such insights have been credited with contributing to breakthroughs in fields such as chemistry, engineering, and computing (Gandy et al., 2022; Sternberg & Davidson, 1995) and appear central to therapeutic outcomes in psychedelic-assisted treatment (Tulver et al., 2023; Kugel et al., 2024). These experiences can lead to lasting cognitive and behavioural changes, as captured by tools such as the Psychological Insight Questionnaire (PIQ) and Psychological Insight Scale (PIS) (Davis et al., 2021; Peill et al., 2022). Although insights are often accurate (Salvi et al., 2016), they can also be illusory and sometimes promote false beliefs (Laukkonen et al., 2020, 2022, 2023; Grimmer et al., 2022, 2023; Tulver et al., 2023; Mason et al., 2021; Timmermann, 2022). This raises the question of whether psychedelic-induced insights are always adaptive and factual. A recent theoretical framework titled *False Insights and Beliefs Under pSychedelics* (FIBUS: McGovern et al., 2024) explicitly argues that reducing the constraining power of prior beliefs during psychedelics increases vulnerability to false insights and the adoption of false beliefs.

**Our Study**

Here we examine cognitive measures related to creativity—divergent thinking, cognitive reflection, and insight—by comparing psychedelic users with non-users in a large sample. Based on previous research, despite inconsistent findings, we hypothesised that past psychedelic use would be associated with higher divergent thinking scores on the DAT. Given literature linking psychedelics to psychological flexibility, we also expected psychedelic users to show higher CRT scores in the long term. Furthermore, due to the well-documented proliferation of insight experiences during psychedelic states, we hypothesised that users would report more insight accompanying CRT responses. However, because subjective feelings of insight do not always translate into accurate problem-solving (McGovern et al., 2024), we did not predict a specific direction of difference in insight accuracy between users and non-users.

**Methods**

*Study Design*

This study was conducted as an optional part of the Great British Intelligence Test January 2022 follow-up, in which participants had the option of completing an additional battery of cognitive tests measuring creativity and cognitive reflection (Hampshire et al., 2021). The study complied with the ethical principles outlined in the Declaration of Helsinki (1975, as revised in 2008). Ethical approval was granted by the Imperial College Research Ethics Committee (17IC4009). All participants provided electronic informed consent before completing the survey. The present study was pre-registered on the Open Science Framework (OSF) on June 23, 2023, and is available at https://osf.io/w9fm7/.

*Participants & Procedure*

Participants were recruited via the Great British Intelligence Test hosted on The Cognitron platform (www.cognitron.co.uk) and led by Professor Adam Hampshire at Imperial College London (Hampshire et al., 2021), with recruitment promoted by BBC media outlets. Advertising occurred in two waves, December 2019 to January 2020 and May 2020, although the website remained open throughout this period, allowing substantial ongoing engagement.

Of the 243,875 individuals who completed the test during this period, 95,441 consented to be recontacted. Follow-up data were collected in January 2022 and included questions about drug use as well as additional cognitive assessments designed to measure creativity. These assessments included the Divergent Association Task (DAT) and the Cognitive Reflection Task (CRT). In total, follow-up data were obtained from 8,917 observations.

*Psychedelic and other Drug Use*

Participants were asked about their recreational drug use during 2019 (pre-pandemic) and 2020–2022 (during the pandemic), covering a range of substances including six classic psychedelics: ayahuasca, magic mushrooms, 5-MeO-DMT, DMT, LSD, and mescaline. Responses across the two time periods were combined, and reports of these six substances were consolidated into a single *psychedelics* category. Participants were then categorised into three groups: *No drug use, Drug use but not psychedelics,* and *Psychedelic use.*

*Cognitive Measures*

Divergent thinking was assessed using the Divergent Association Task (DAT), a brief, reliable, and objective measure of divergent thinking (Olson et al., 2021). Participants listed 10 unrelated nouns, which were analysed for semantic distance using open-access software (www.datcreativity.com), with scores above 100 considered indicative of high divergent thinking. Scores were computed only when participants provided at least seven correctly spelled nouns; entries with fewer words were excluded. Scores below 50 were also excluded, as they typically reflected semantically similar words, suggesting the participant had not understood the task. In total, 742 entries were excluded, yielding a final DAT sample of 5,163 participants.

Cognitive reflection was measured using a four-item Cognitive Reflection Task (CRT), comprising the original three items (Frederick, 2005) and an additional item from Primi et al. (2016). Responses were aggregated and coded to account for variations in answer formats. Participants who left three or more questions unanswered were excluded. In total, 202 cases were removed, resulting in a final CRT sample of 5,703 participants. The CRT questions and correct answers are presented in Appendix A.

Insight was measured by asking participants to report whether solving each CRT item was accompanied by an *Aha!* moment. Appendix B contains the definition of an *Aha!* moment provided to participants. Because insight reports were tied to CRT responses, the sample size for the number of insights matched the CRT sample (N = 5,703). For each participant, we calculated the number of insights, as well as *insight accuracy*, defined as the proportion of insights that accompanied correctly solved items. Not all participants reported any insights; therefore, analyses of insight accuracy were restricted to those who reported at least one insight (N = 2,725).

### *Design & Statistical Analysis*

All analyses were conducted using a combination of Jamovi (version 2.3.26.0) for data cleaning and initial screening, and Python (version 3.14.0) for the main statistical modelling.

### *Data Screening and Assumption Testing*

Data screening involved evaluating normality, linearity, homogeneity of variance, normality of residuals, and homoscedasticity. Full information is presented in Appendix C. Each cognitive outcome (DAT, CRT, Insight Count, Insight Accuracy) was first regressed on age (in decades), sex, and education using ordinary least squares (OLS). The residuals from these models were then transformed using a rank-based inverse normal transformation to normality. Visual inspection of diagnostic plots indicated that the transformed residuals met the assumptions required for one-way ANOVA. Complete plots and descriptive notes are provided in Appendix C.

### *Main Analyses*

Because demographic variables differed between drug-use groups and could plausibly influence cognitive performance, each cognitive outcome (DAT, CRT, Insight Count, Insight Accuracy) was first adjusted for age (in decades), sex, and education level using ordinary least squares (OLS) regression. The residuals from these models were taken as covariate-adjusted scores. These residuals were then transformed using a rank-based inverse normal transformation to normality. Each outcome variable was analysed separately using a one-way ANOVA, with drug-use group (*No drug use, Drug use, Psychedelic use*) entered as the between-participants factor. When the omnibus test was significant, Tukey's Honest Significant Difference (HSD) test was used for post-hoc comparisons, controlling the family-wise error rate. All analyses were two-tailed with α set at .05. DAT, CRT, Insight Count, and Insight Accuracy were treated as distinct dependent variables.

## Results

### *Demographics*

The final dataset consisted of 5,905 participants after excluding 2,032 individuals who did not complete the drug-use questions, 576 who reported drug use but declined to provide further details, and 404 duplicate responses. Descriptive statistics and demographic characteristics of the retained sample are presented in Tables 1 and 2.

**Table 1.**
*Participant breakdown*

| Category | N | % |
|---|---|---|
| Total Participants (Before Exclusions) | 8,917 | 100.00 |
| Excluded Participants | 3,012 | 33.77 |
|    Did not complete drug information | 2,032 | 22.79 |
|    Used drugs but declined further details | 576 | 6.46 |
|    Duplicated responses | 404 | 4.53 |
| Retained Participants (Final Dataset) | 5,905 | 66.23 |

**Table 2.**
*Demographics of Retained Participants (N = 5,905)*

| Category | N | % |
|---|---|---|
| Gender | | |
|    Female | 3,566 | 60.39 |
|    Male | 2,309 | 39.10 |
|    Other | 30 | 0.51 |
| Age (*M*, *SD*) | 51.13 (15.53) | — |
| Education Level | | |
|    Did not complete high school | 90 | 1.52 |
|    Completed high school | 1,821 | 30.84 |
|    Obtained a degree | 3,679 | 62.30 |
|    Completed a PhD | 315 | 5.33 |

*Descriptive Statistics*

Drug use categories and their socio-demographics are shown in Table 3.

**Table 3**
*Participant drug use categories and sociodemographics*

| age/Drug group | n | % | age (M) | age (SD) |
|---|---|---|---|---|
| No drug use | 4289 | 72.63 | 52.24 | 15.73 |
| Drug use but not psychedelics | 1520 | 25.74 | 48.92 | 14.42 |
| Psychedelic use | 96 | 1.63 | 37.03 | 13.01 |

| Gender/Drug group | Male | Female | Other |
|---|---|---|---|
| No drug use | 1504 (65.14%) | 2764 (77.51%) | 21 (70%) |
| Drug use but not psychedelics | 752 (32.57%) | 760 (21.31%) | 8 (26.67%) |
| Psychedelic use | 53 (2.30%) | 42 (1.18%) | 1 (3.33%) |

| Education/Drug group | *No high school* | *High school* | *University* | *PHD* |
|---|---|---|---|---|
| No drug use | 73 (80.22%) | 1370 (74.71%) | 2618 (70.89%) | 228 (70.58%) |
| Drug use but not psychedelics | 16 (17.58%) | 423 (23.07%) | 998 (27.02%) | 83 (25.69%) |
| Psychedelic use | 1 (1.10%) | 24 (1.13%) | 63 (1.71%) | 4 (1.24%) |

A Levene's test for homogeneity of variances indicated a significant violation of the equal-variance assumption for age across the three drug-use groups, $F(2, 5902) = 7.27$, $p < .001$. Accordingly, a Welch's ANOVA—robust to unequal variances—was used to examine age differences among the *no drug use*, *drug use but not psychedelics*, and *psychedelic use* groups. The analysis showed a statistically significant effect of group on age, $F(2, 256.09) = 84.03$, $p < .001$, indicating that mean age differed across groups. Post-hoc inspection showed that psychedelic users were younger than participants in the other two groups.

Levene's test also indicated a violation of homogeneity for gender proportions across drug-use groups, $F(2, 5902) = 62.88$, $p < .001$. Because this violates assumptions for parametric comparisons, a chi-square test of independence was conducted to examine the association between gender and drug-use group. The test yielded a statistically significant association, $\chi^2(4, N = 5,905) = 109.87$, $p < .001$. Standardised residuals indicated that females were overrepresented in the *no drug* use and *drug use but not psychedelics* groups and underrepresented in the *psychedelic use* group, whereas males were more likely to report psychedelic use relative to other genders.

A chi-square test of independence was also conducted to assess the association between education level and drug-use group. Results showed a statistically significant association, $\chi^2(6, N = 5,905) = 13.81$, $p = .032$. Although the effect was small, the pattern suggested that higher education was more common among individuals reporting drug use, particularly those reporting psychedelic use. Of the 96 participants who used psychedelics, 90 (93.75%) also reported other drug use, and 81 (84.38%) had used cannabis.

*Divergent Association, Cognitive Reflection, and Insight*

For each of our measures of DAT, CRT, number of insights, and insight accuracy, we compared scores among psychedelic users, non–drug users, and users of drugs other than psychedelics. The mean DAT score for the analysed sample ($N = 5,163$) was 78.10 (*SD* = 7.39), 95% *CI* [77.90, 78.31]. The mean Cognitive Reflection Task (CRT) score for the sample ($N = 5,777$) was 2.24 (*SD* = 1.21), 95% *CI* [2.21, 2.28]. For insight count, the mean number of insights accompanying the CRT items ($N = 5,703$) was 0.84 (SD = 1.08), 95% *CI* [0.81, 0.87]. This sample was slightly smaller than the CRT sample, as not all participants responded to the insight questions.

For insight accuracy, the mean score ($N = 2,731$) was 75.77% (*SD* = 36.74), 95% *CI* [74.39, 77.15]. This subsample was smaller because participants who reported no insights could not be given an accuracy score. Descriptive statistics for all measures are presented in Table 4.

**Table 4**
*Descriptive Statistics for each measurement by Drug Group*

| Measure/Drug group | *n* | *M* | 95% CI | *SD* |
|---|---|---|---|---|
| Divergent Association Task (DAT) | | | | |
| No drug use | 3718 | 77.62 | [77.38, 77.86] | 7.40 |
| Drug use but not psychedelics | 1357 | 79.17 | [78.79, 79.56] | 7.23 |
| Psychedelic use | 88 | 81.99 | [80.60, 83.38] | 6.60 |
| Cognitive Reflection Task (CRT) | | | | |
| No drug use | 4197 | 2.19 | [2.15, 2.23] | 1.21 |
| Drug use but not psychedelics | 1488 | 2.36 | [2.30, 2.42] | 1.22 |
| Psychedelic use | 92 | 2.79 | [2.57, 3.01] | 1.06 |
| Number of Insights | | | | |
| No drug use | 4137 | 0.83 | [0.80, 0.86] | 1.09 |
| Drug use but not psychedelics | 1474 | 0.86 | [0.80, 0.91] | 1.08 |
| Psychedelic use | 92 | 0.77 | [0.55, 0.99] | 1.05 |
| Insight Accuracy | | | | |
| No drug use | 1954 | 74.86 | [73.22, 76.51] | 37.12 |
| Drug use but not psychedelics | 737 | 77.83 | [75.23, 80.43] | 35.94 |
| Psychedelic use | 40 | 82.29 | [72.37, 92.21] | 31.02 |

*Divergent Thinking*

Psychedelic users in our sample leaned disproportionately toward male, young, and more-educated participants. To account for demographic differences between groups, DAT scores were first adjusted for age (in decades), sex, and education using linear regression, and the resulting residuals were transformed using a rank-based inverse normal transformation. A one-way ANOVA on the adjusted scores showed a significant effect of drug group, $F(2, 5160) = 22.98$, $p < .001$. Tukey HSD post-hoc tests indicated a clear stepwise pattern:

*Drug use* scored higher than the *No drug use* group
(mean difference = 0.177, $p < .001$)
*Psychedelic use* scored higher than the *No drug use* group
(mean difference = 0.454, $p < .001$)
*Psychedelic use* also scored higher than *Drug use but not psychedelics*
(mean difference = 0.277, $p = .031$)

Overall, divergent thinking scores increased progressively from *No drug use* → *Drug use but not psychedelics* → *Psychedelic use*.

**Figure 1.**
*Divergent Association Task (DAT) scores across drug groups*

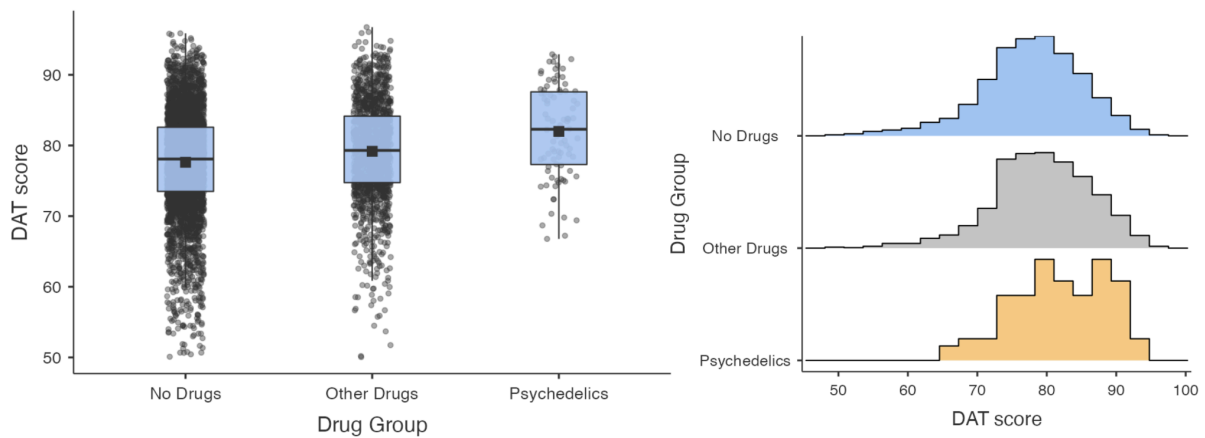

*Note.* Mean scores and box plots (left), frequency distribution (right). Participants in the "other drug" group used drugs but not psychedelics.

### *Cognitive Reflection*

We compared the CRT scores of psychedelic users with non-drug users and with users of drugs without psychedelics. CRT scores were adjusted for age, sex, and education using linear regression and then transformed using a rank-based inverse normal transformation. A one-way ANOVA on the adjusted scores showed no significant effect of drug group, $F(2, 5774) = 1.71$, $p = .18$. Tukey HSD post-hoc comparisons confirmed no significant differences between:

*No drug use* vs. *Drug use but not psychedelics*: $p = .917$
*No drug use* vs. *Psychedelic use*: $p = .160$
*Drug use but not psychedelics* vs. *Psychedelic use*: $p = .211$
CRT performance did not differ meaningfully between the groups.

**Figure 2.**
Cognitive Reflection Task (CRT) scores across drug groups

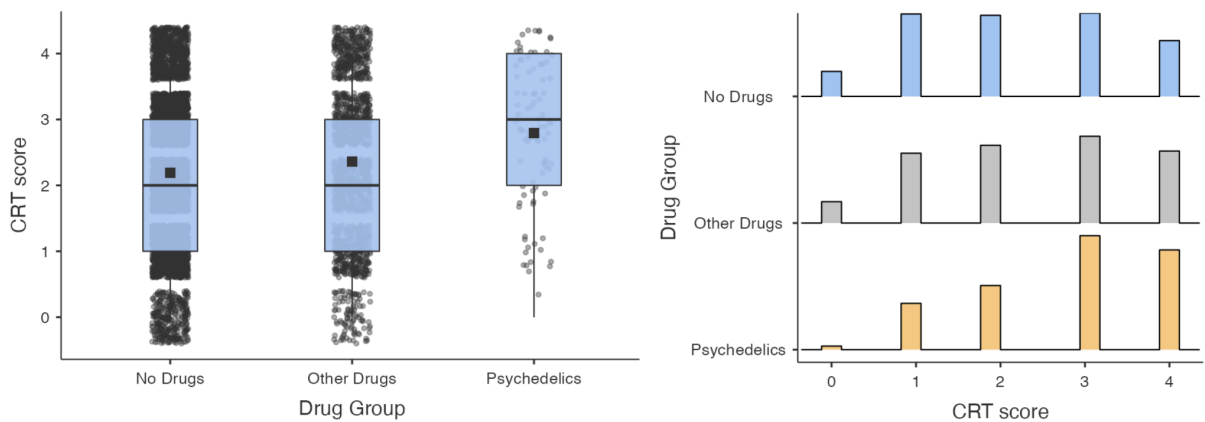

*Note.* Mean scores and box plots (left), frequency distribution (right). Participants in the "other drug" group used drugs but not psychedelics.

### *Insight*

We compared the insight scores of psychedelic users with non-drug users and with users of drugs but not psychedelics. To adjust for demographic differences, the number of insights was residualised for age (in decades), sex, and education, and then transformed using a rank-based inverse normal transformation. A one-way ANOVA on the adjusted scores showed a significant effect of drug group, $F(2, 5700) = 5.72$, $p = .003$. Tukey HSD post-hoc tests showed that:

*Drug use but not psychedelics* reported slightly more insights than the *No drug use*
(mean difference = 0.1015, *p* = .002)
*Psychedelic use* did not differ from the *No drug use*
(mean difference = 0.0623, *p* = .823)
*Psychedelic use* did not differ from *Drug use but not psychedelics*
(mean difference = –0.0392, *p* = .929)

Thus, although drug users reported marginally more insights than non-users after demographic adjustment, psychedelic use was not associated with additional insight experiences.

**Figure 3.**
*Mean number of insights during the Cognitive Reflection Task (CRT) across drug groups*

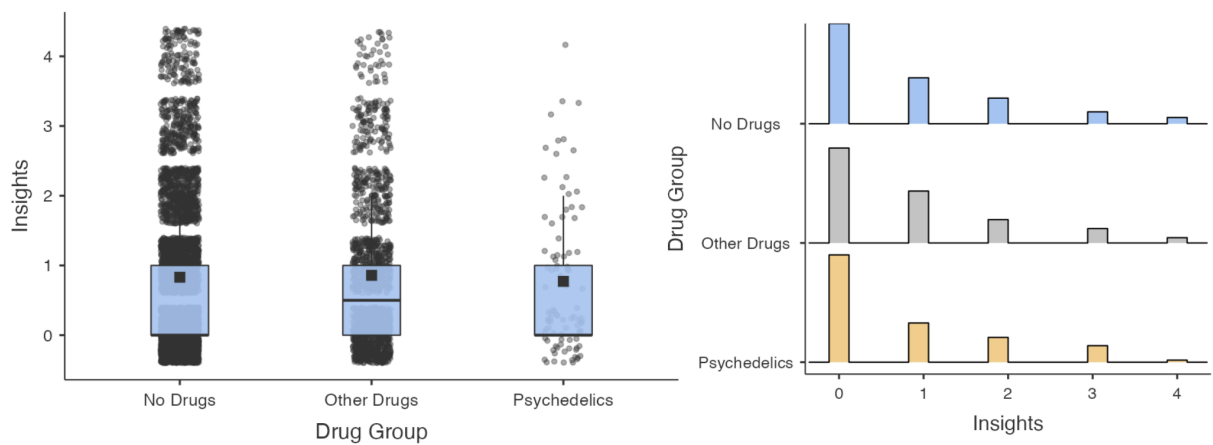

*Note.* Mean scores and box plots (left), frequency distribution (right). Participants in the "other drug" group used drugs but not psychedelics.

*Insight Accuracy*

We then compared the insight accuracy scores of psychedelic users with non-drug users and with users of other drugs besides psychedelics. Insight accuracy scores were adjusted for age, sex, and education using linear regression, followed by a rank-based inverse normal transformation. A one-way ANOVA revealed no significant differences between drug-use groups, $F(2, 2728) = 0.98$, $p = .38$. Tukey HSD post-hoc tests confirmed no significant pairwise differences:

*No drug use* vs. *Drug use but not psychedelics*: *p* = .607
*No drug use* vs. *Psychedelic use*: *p* = .518
*Drug use but not psychedelics* vs. *Psychedelic use*: *p* = .689
Hence, insight accuracy did not differ between conditions.

*Polydrug Use: Exploratory*

Mean scores on the DAT and CRT were also compared across different drug user groups. Since many participants used multiple substances, they were included in multiple categories. For example, in our sample of 98 psychedelic users, 81 also used cannabis, meaning they were counted in both the psychedelic and cannabis groups, as well as any other relevant categories. This overlap underscores the challenge of isolating substance-specific effects on creativity-related cognition. In light of potential confounds associated with polydrug use, Figure 4 illustrates the general trends for each drug group. Heroin was excluded due to the small sample size ($N = 6$).

**Figure 4.**
*Mean scores on the DAT and CRT across different drug categories*

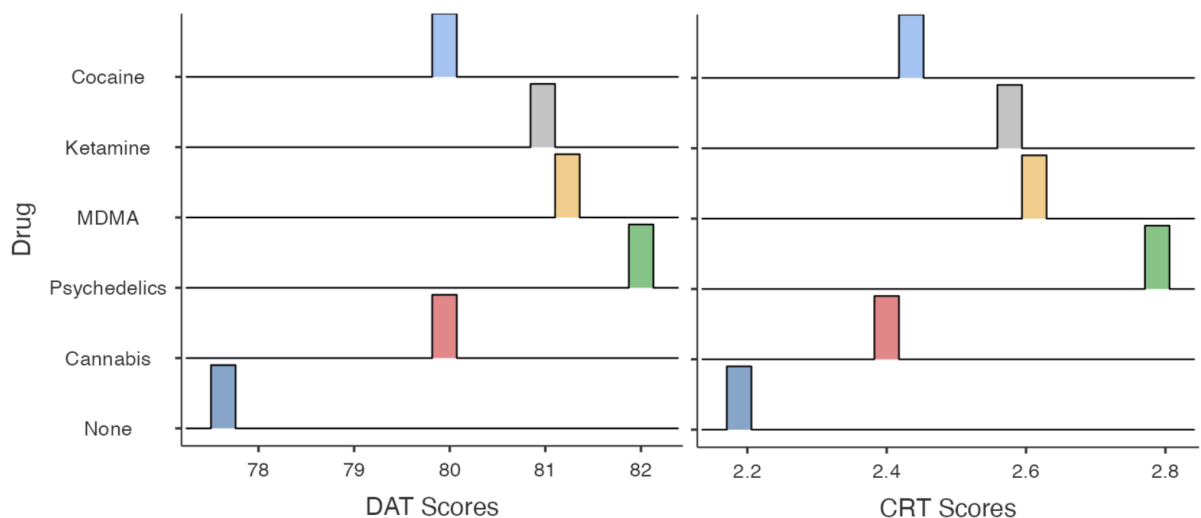

*Note.* Mean scores for DAT (left) and CRT (right) by drug group. *N* for each drug group is as follows, noting that polydrug users will appear in more than one drug group: DAT: 146 (cocaine), 48 (ketamine), 104 (MDMA), 88 (psychedelics), 465 (cannabis); CRT: 156 (cocaine), 48 (ketamine), 110 (MDMA), 92 (psychedelics), 505 (cannabis).

**Discussion**

The present study examined whether past psychedelic use was associated with different components of creativity—specifically divergent thinking, cognitive reflection, and insight. We found that individuals with a history of psychedelic use scored significantly higher on divergent thinking than both non–drug users and users of drugs other than psychedelics. In contrast, there were no significant group differences in cognitive reflection, number of insights, or insight accuracy. Taken together, these results suggest a more selective association between psychedelic use and divergent thinking, one of the core processes involved in creative cognition.

These findings are broadly consistent with anecdotal reports and theoretical proposals that psychedelics may influence creative thinking (Gandy et al., 2022; Wießner et al., 2022). The higher divergent thinking scores among psychedelic users support the idea that psychedelics may enhance the fluency or flexibility with which people generate novel ideas. This mirrors early work such as Harman et al. (1966), who observed improved performance on alternative use tasks after mescaline administration, although their methods—and those of similar early studies—had important limitations (Baggott, 2015). Of course, the present findings are correlational, and causation cannot be inferred, a point addressed further below.

The absence of significant differences in cognitive reflection and insight measures suggests that psychedelics may relate more strongly to the generative aspects of creativity than to evaluative or metacognitive aspects. Cognitive reflection—the ability to override intuitive responses in favour of more deliberate reasoning (Frederick, 2005; Erceg et al., 2020)—showed a slight trend toward higher scores for psychedelic users, but this did not reach significance. This pattern indicates that psychedelics may influence the production of novel ideas (divergent thinking) without necessarily affecting the capacity to evaluate or refine those ideas. This interpretation aligns with work suggesting that psychedelic effects tend to impact the generation of new content more than the metacognitive processes involved in assessing that content (Bayne & Carter, 2018; Preller & Vollenweider, 2018).

*Mechanisms Underlying Enhanced Divergent Thinking*

Although our findings are correlational, several theoretical accounts support the possibility of a more meaningful, potentially causal relationship. The association between psychedelic use and increased divergent thinking may be influenced by known neurobiological mechanisms. Psychedelics act as 5-HT2A receptor agonists, leading to increased neural entropy and the breakdown of higher-order brain networks such as the default mode and frontoparietal networks, which support internally and externally oriented cognition (Carhart-Harris & Friston, 2019; Dixon et al., 2018; Timmermann et al., 2019; Yeshurun et al., 2021). This disruption may loosen rigid thought patterns and allow more flexible, novel associations to emerge (Carhart-Harris et al., 2014; McGovern et al., 2024).

Alongside this breakdown, these same network hubs often show increased functional connectivity with the rest of the brain (Tagliazucchi et al., 2016; Timmermann et al., 2023), suggesting that cognitive resources associated with higher-order cognition may be redistributed during the psychedelic state. Such changes could support increased divergent thinking and may persist to some extent after acute effects have subsided. Psychedelics have also been proposed to promote synaptic plasticity and neurogenesis (Bouso et al., 2015; Liao, 2024; Kim et al., 2023), which could contribute to longer-term shifts in cognitive processes relevant to creativity.

The REBUS (Relaxed Beliefs Under Psychedelics) model further proposes that psychedelics relax high-level priors, allowing for a greater bottom-up flow of information and encouraging exploration of new cognitive pathways (Carhart-Harris & Friston, 2019).

This framework is consistent with our findings, insofar as psychedelic use may enhance divergent thinking by promoting cognitive flexibility through underlying neurophysiological changes. However, as noted earlier, divergent thinking does not guarantee accuracy or fidelity, and the emergence of new cognitive pathways does not imply that they are necessarily adaptive or beneficial (McGovern et al., 2024; Laukkonen et al., 2023).

*Alternative Explanations*

While our findings indicate a relationship between past psychedelic use and divergent thinking, alternative explanations should be considered. It is possible that individuals who are inherently more creative are also more inclined to experiment with psychedelics (Erritzoe et al., 2019; McCrae, 1987). In that case, the observed associations may reflect pre-existing differences not fully accounted for in our model rather than effects of psychedelic use. Notably, higher divergent thinking scores among users of any substance suggest that drug use more broadly may relate to creativity. However, the significantly higher scores among psychedelic users compared with all other drug groups point to a potentially unique association for this drug class. This distinction matters, as it suggests that the enhancement is not simply a byproduct of general substance use but may relate specifically to the pharmacological properties of psychedelics.

*Strengths and Limitations*

There are several strengths associated with this study. It represents one of the few investigations examining the effects of naturalistic psychedelic use without a psychedelic specific recruitment bias. The data were not collected in response to advertisement materials specifically recruiting drug users, either during initial recruitment or prior to follow-up, and the study was never advertised on social media channels related to psychedelics or drug use. In addition, the sample includes a diverse population, with a large control group who had never used drugs yet completed the same measures over the same period.

Several limitations also warrant consideration. First, the cross-sectional design limits our ability to infer causality. Prospective longitudinal studies are necessary to determine whether psychedelic use leads to sustained enhancements in divergent thinking. Although we controlled for demographic variables such as age, gender, and education, other confounding factors—such as personality traits like openness to experience (McCrae, 1987) or socio-cultural influences—may influence both psychedelic use and creativity. Previous research has shown that openness to experience is associated with both creativity and psychedelic use (Erritzoe et al., 2019). Our measures of creativity also focused primarily on divergent thinking as assessed by the DAT. While the DAT is a reliable and objective measure (Olson et al., 2021), creativity is multifaceted, and aspects such as convergent thinking, artistic creativity, or real-world creative achievements were not captured. Future research should therefore include a broader range of creativity measures to more fully encompass the construct.

Finally, we did not collect data on frequency of psychedelic use or commonly used doses. This is particularly relevant given the growing popularity of microdosing, for which several placebo-controlled studies report only mild or null effects on wellbeing and cognition relative to higher doses of psychedelics (Murphy et al., 2023; Szigeti et al., 2021).

*Implications and Future Directions*

Our findings contribute to the growing body of literature exploring the cognitive effects of psychedelics beyond their therapeutic applications. There is increasing public and scientific interest in the potential of psychedelics for cognitive enhancement (Bălăeț, 2025). The observed association between past psychedelic use and enhanced divergent thinking suggests that these substances may have relevance in domains that prioritise creativity and innovation (Damer, 2023). This raises intriguing possibilities for future research into cognitive enhancement and for developing interventions aimed at supporting creative thinking.

Clearly, future studies should work to establish causality through well-controlled experimental designs. Randomised controlled trials that administer psychedelics and assess creativity over time would provide more definitive evidence. Incorporating computerised cognitive testing may be particularly useful for evaluating creativity across a range of contexts (Bălăeț, 2022). In parallel, investigating underlying neural mechanisms through neuroimaging could help clarify how psychedelics influence creative cognition. Given the prevalence of polydrug use (Bălăeț et al., 2023; Bălăeț et al., 2025; Bălăeț et al., 2025), employing advanced statistical techniques to analyse drug-use data may also be warranted.

**Conclusion**

Our study provides early evidence that past naturalistic psychedelic use is associated with higher levels of divergent thinking, a key component of creativity. This association remained even after controlling for other drug use and demographic variables. The findings align with theoretical accounts and prior research suggesting that psychedelics may enhance aspects of creative thinking, though not necessarily insightfulness.

**Acknowledgements:** We would like to thank Professor Adam Hampshire and the Great British Intelligence Test participants. ChatGPT (OpenAI) was used solely to assist with language editing and refinement of text after the initial manuscript was written. No generative AI tools were used to create ideas, interpretations, analyses, or substantive content. All text was reviewed, verified, and approved by the authors, who take full responsibility for the final manuscript.

**Authorship contribution:** Initial Idea: CT; Conceptualisation: REL, CT, MB, GP; Methodology: REL, CT, MB Data Processing: MB; Data Collection & Software: WT, PJH, MB; Investigation: GP, MB; Visualisation: GP, REL; Supervision: REL, MB, CT; Writing – Original Draft: GP, REL; Writing – Review & Editing: GP, MB, WT, PJH, CT, REL

**Appendices**

*Appendix A*

The four cognitive reflection questions are listed below:
1. A bat and a ball cost $1.10 in total. The bat costs $1.00 more than the ball. How much does the ball cost?
2. If it takes 5 machines 5 minutes to make 5 widgets, how long would it take 100 machines to make 100 widgets?
3. In a lake, there is a patch of lily pads. Every day, the patch doubles in size. If it takes 48 days for the patch to cover the entire lake, how long would it take for the patch to cover half of the lake?
4. If three elves can wrap three toys in 1 hour, how many elves are needed to wrap six toys in 2 hours?

The intuitive response to the first question is 10 cents, however the correct answer is 5 cents. The intuitive response to question two is 100, however the correct answer is 5. The intuitive response to question three is 24, however the correct answer is 47. The intuitive response to question four is 6, however the correct answer is 3.

*Appendix B*

Explanation of "Aha!" Moment Given to Participants:

The explanation of "Aha!" moment given to participants was as follows:

After you decide whether the claim is true, you will be asked whether or not you experienced an "Aha!" moment at any point in the trial. Almost everyone has experienced an Aha! moment in the past. Many people report Aha! moments while having a shower, or just before falling asleep. Try to recall an Aha! experience that you've had, and try to remember how it felt.

When completing the task, try to pay attention to when Aha! moments occur. When an Aha! moment occurs, it is as if the solution to the problem suddenly pops into your mind, like a lightbulb turning on. You might experience surprise, you might feel relief, and you might feel a light sense of happiness and ease. You can think of this experience as a miniature 'Eureka moment'. You might even feel an internal sense of "Aha!," or you might think to yourself, "of course!," "that was so obvious". Not experiencing an Aha! moment might feel like nothing much at all. You might simply think about the problem, and then gradually work out the solution.

*Appendix C*

Prior to conducting the primary analyses, each cognitive outcome (Divergent Association Task; Cognitive Reflection Task; Number of Insights; Insight Accuracy) was regressed on age (in decades), sex, and education using ordinary least squares (OLS). The residuals from these models were then subjected to a rank-based inverse normal transformation to approximate a normal distribution.

Figures C1-C4 display the distributions of the transformed residuals for each outcome variable. Visual inspection indicated that the transformed values closely approximated

normality across all measures. Slight departures from normality were evident for the insight variables due to their restricted raw score ranges, but these deviations were minimal and unlikely to influence the robustness of the subsequent one-way ANOVAs.

Because the ANOVAs were performed on residualised and normalised scores, ANCOVA-specific assumptions (e.g., homogeneity of regression slopes) were not applicable. Homogeneity of variance was evaluated via visual inspection of residual plots and judged to be acceptable for all models. Overall, the transformed residuals met the assumptions required for the one-way ANOVAs reported in the Results section.

Figure C1-C4. Transformed residuals for each outcome variable

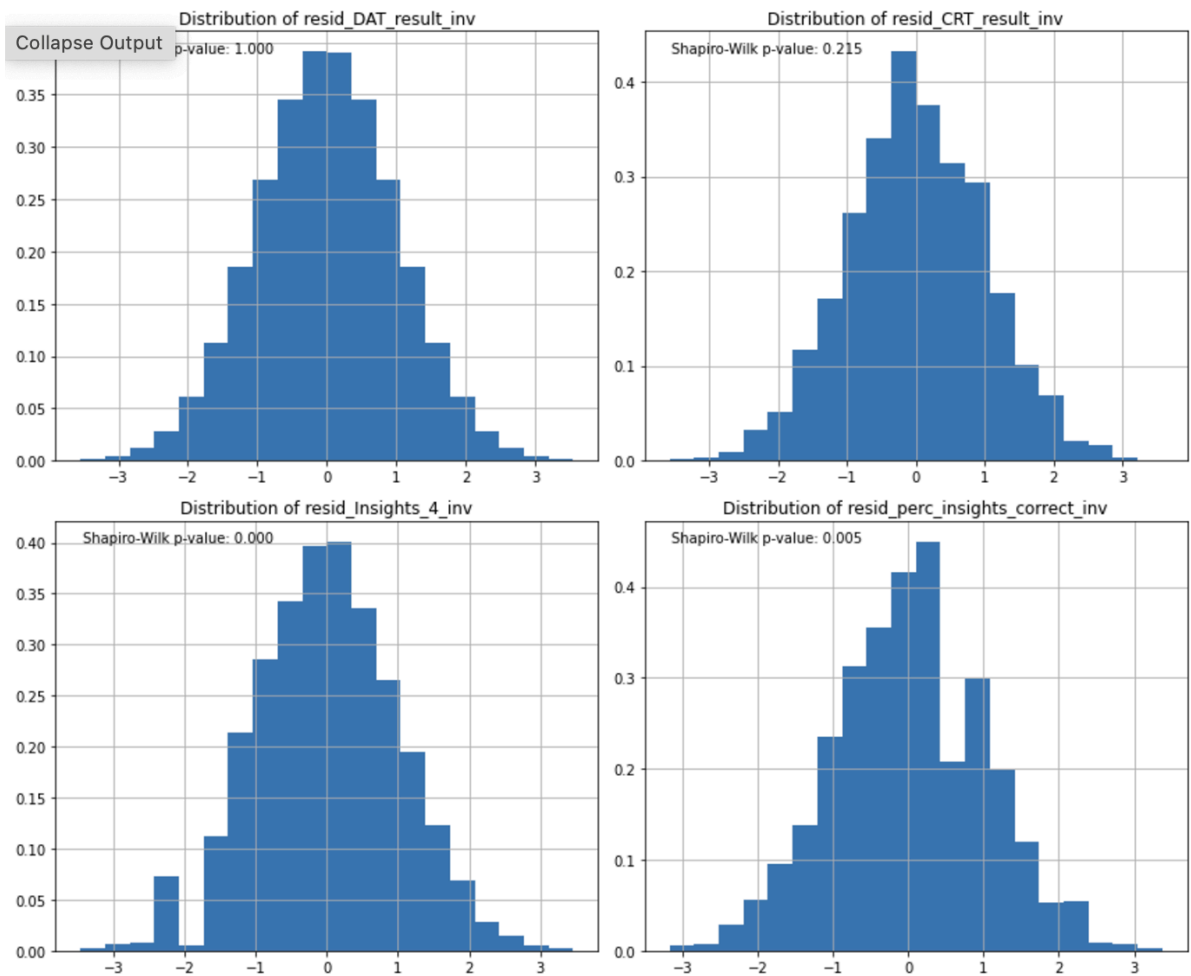